\documentclass[a4paper,fleqn,usenatbib]{mnras}

\usepackage{newtxtext,newtxmath}
\usepackage[T1]{fontenc}
\usepackage{ae,aecompl}

\usepackage{graphicx}	% Including figure files
\usepackage{amsmath}	% Advanced maths commands
\usepackage{amssymb}	% Extra maths symbols

\newcommand{\Equ}[1]{Eq.~(\ref{eq:#1})}

\newcommand{\se}[1]{\S\ref{sec:#1}}

\newcommand{\Fig}[1]{Figure~\ref{fig:#1}}

\newcommand{\tab}[1]{Table~\ref{tab:#1}}
\newcommand{\be}{\begin{equation}}
\newcommand{\ee}{\end{equation}}
\newcommand{\bea}{\begin{eqnarray}}
\newcommand{\eea}{\end{eqnarray}}

\newcommand{\msun}{{\rm M}_\odot}

\newcommand{\ifm}[1]{\relax\ifmmode#1\else$\mathsurround=0pt #1$\fi}
\newcommand{\kms}{\ifmmode\,{\rm km}\,{\rm s}^{-1}\else km$\,$s$^{-1}$\fi}
\newcommand{\hmpc}{\,\ifm{h^{-1}}{\rm Mpc}}

\newcommand{\Mpc}{\,{\rm Mpc}}

\newcommand{\ltsima}{$\; \buildrel < \over \sim \;$}
\newcommand{\lsim}{\lower.5ex\hbox{\ltsima}}
\newcommand{\gtsima}{$\; \buildrel > \over \sim \;$}
\newcommand{\gsim}{\lower.5ex\hbox{\gtsima}}

\def\la{\langle}

\def\omm{\Omega_{\rm m}}

\def\omb{\Omega_{\rm b}}
\def\sy{\,M_\odot\, {\rm yr}^{-1}}

\def\cmc{\,{\rm cm}^{-3}}
\def\cms{\,{\rm cm}^{-2}}

\def\Mv{M_{\rm vir}}

\def\Ms{M_*}

\usepackage{color}

\title[FirstLight]{Introducing the FirstLight Project: UV luminosity function and scaling relations of primeval galaxies}

\author[Ceverino et al.]{
Daniel Ceverino,$^{1}$\thanks{E-mail: ceverino@uni-heidelberg.de}
Simon C. O. Glover,$^{1}$ Ralf S. Klessen$^{1}$
\\
$^{1}$Universität Heidelberg, Zentrum für Astronomie, Institut für Theoretische Astrophysik, Albert-Ueberle-Str. 2, 69120 Heidelberg, Germany\\
}

\date{Accepted XXX. Received YYY; in original form ZZZ}

\pubyear{2017}

\begin{document}
\label{firstpage}
\pagerange{\pageref{firstpage}--\pageref{lastpage}}
\maketitle

\begin{abstract}

We introduce the FirstLight project that aims to generate a large database of high-resolution, zoom-in simulations of galaxy formation around the epoch of reionisation ($z\geq6$). 
The first results of this program agree well with recent observational constraints at $z=6-8$, including the UV luminosity function and galaxy stellar mass function, as well as the scaling relationships between halo mass, stellar mass, and UV magnitude.
The UV luminosity function starts to flatten below $M_{UV}>-14$ due to stellar feedback 
in halos with maximum circular velocities of $V=30-40 \kms$.
The power-law slope of the luminosity function evolves rapidly with redshift, reaching a value of $\alpha\simeq-2.5$ at $z=10$.
On the other hand, the galaxy stellar mass function evolves slowly with time between $z=8-10$, in particular at the low-mass end. 
\end{abstract}


%%%%%%%%%%%%%%%%% BODY OF PAPER %%%%%%%%%%%%%%%%%%

%%%%%%%%%%%%%%%%%%%%%%%%%%%%%%%%%%%%%%%%%%%%%%%%%%
\section{Introduction}
%%%%%%%%%%%%%%%%%%%%%%%%%%%%%%%%%%%%%%%%%%%%%%%%%%

The formation of the first stars and galaxies marks the beginning of the cosmic dawn,
when stellar light spreads across the cosmos.
The first light of these primeval galaxies reshapes the global properties of the Universe during the epoch of reionisation.
However, very little is known about the properties of galaxies during the first billion years of the Universe.

One of the basic properties during the reionisation epoch is the abundance of galaxies as a function of their UV luminosities. 
The shape and evolution of the UV luminosity function (UVLF) gives insight into the efficiency of star formation as a function of halo mass and time.
This is crucial for assessing the importance of galaxies for the reionisation of the Universe.
Surveys from HST fields have yielded a large population of galaxies between redshifts $z=4$ to $z=10$ \citep{Bouwens04, Finkelstein12, Oesch13, Bouwens15}.
This allows the accurate determination of the UVLF at $z\leq8$ and its evolution.
Observations indicate a decrease in the normalisation with increasing redshift, as well as a steepening of the low-luminosity slope.
However, large uncertainties remain at higher redshifts ($z\ge10$), as well as at the high and low luminosity ends ($M_{UV}<-22$ and $M_{UV}>-16$), due to the low number of observed galaxies.
Future deep surveys using the James Webb Space telescope (JWST) will significantly improve this situation.
Meanwhile, theoretical predictions are crucial for the design of these future surveys.

Any theory of galaxy formation should predict the right correlations between three basic galaxy properties: the mass of a halo (or its virial mass, $\Mv$), the stellar mass of the galaxy at the centre of this halo ($\Ms$), and the star formation rate (SFR) or the equivalent UV magnitude ($M_{UV}$) within the galaxy.
Observations are starting to constrain these scaling relations at high-z \citep{Stark13, Duncan14, Song16, Stefanon16}. 
Taking into account the nebular emission lines, current determinations of the $M_{UV}-\Ms$ relation at different redshifts show a weak evolution with time.
However, the uncertainties are still large.
The galaxy stellar mass function has also been measured at $4<z<8$ \citep{Song16, Stefanon16} and these observations report a steep low-mass end, although its evolution is a matter of debate.

%\adb{specially if dusty galaxies are missing from current UV-selected samples.}

Many theoretical studies predict a flattening or turn-over in the UVLF at low luminosities \citep{Jaacks13, Dayal14, BoylanKolchin15, Gnedin16}. 
However, a disagreement remains about the details of this flattening. For example, \cite{Gnedin16} and \cite{Liu16} report a break for $M_{UV}>-12$ 
due to a decrease in the efficiency of cooling processes in low-mass halos, 
while \cite{Jaacks13} find a turn-over at $M_{UV}\simeq-15$. 
Other proposed mechanisms include a greater role of radiative feedback that quenches star formation in halos below a given circular velocity of $30-50 \ \kms$ \citep{OShea15, Ocvirk16, Yue16}.  

Due to the importance of the UVLF,
the scaling relations between halo mass, stellar mass and SFR have received less attention, particularly in the regimes where there are observational constraints. Some exceptions include
the $\Ms-$SFR relation \citep{Pawlik17, Cullen17}, and
the $M_{UV}-\Mv$ relation \citep{Finlator17, Liu16},
where simulations tend to overproduce stars in comparison with simple models of abundance matching \citep{Finkelstein15}.

% Motivation:

The FirstLight project is motivated by the need for a large, statistically significant sample of galaxies simulated at very high resolution at the epoch of reionisation ($z\geq6$). 
Current simulations of galaxy formation at these redshifts yield a large population of galaxies in large volumes but the internal properties of their interstellar medium are poorly resolved \citep{Genel14, Pawlik17} and rely heavily on subgrid modeling.
 On the other hand, current zoom-in simulations \citep{Ma15, Fiacconi17, Pallottini17} concentrate all the computational resources in one or just a few galaxies with much higher resolution.
  However, the small sample makes them very sensitive to selection effects and poor statistics. 
 
 The FirstLight project is the largest sample of zoom-in, initial conditions carried out to date that will reach
 typical resolutions of about 10 parsecs in volumes of up to $\sim$60 Mpc.
 Such a large program needs first a validation test.
 We need to make sure that simulated galaxy properties, in particular halo mass, stellar mass and SFR, are consistent with the observed luminosity and galaxy mass function, as well as the scaling relations constrained by observations and other independent methods, such as abundance matching \citep{BehrooziSilk15}.
 
 %Outline:
 
 The outline of this paper is the following. Section 2 describes the full FirstLight sample, as well as the simulations details and the initial conditions of the zoom-in simulations that comprise the test runs (\se{IC}).
 Section 3 describes the results of the FirstLight tests at $z=6$ (\se{six}).
 The following sections are devoted to the evolution of the stellar-to-halo mass relation (\se{SMHM}), the UV luminosity function (\se{evoLF}), and the galaxy stellar mass (\se{evoMs}). Section 7 finishes with the conclusion and final discussions (\se{conclusions}).

%%%%%%%%%%%%%%%%%%%%%%%%%%%%%%%%%%%%%%%%%%%%%%%%%%
\section{The Simulations}
\label{sec:IC}

\subsection{The FirstLight Sample}

The FirstLight project consists of a mass-limited sample of halos with a maximum circular velocity, V, between 50 $\kms$ and 250 $\kms$, selected at $z=5$. The sample covers a halo mass range between a few times $10^9 \ \msun$ and a few times $10^{11} \ \msun$. 
This range excludes massive and rare halos with number densities lower than $\sim10^{-4} h^{-1} \Mpc^{-3}$, 
as well as small halos in which galaxy formation is extremely inefficient.

The sample uses two different sets of cosmological parameters: WMAP5 with $\omm=0.27$, $\omb=0.045$, h=0.7, $\sigma_8=0.82$ \citep{Komatsu09} and  Planck13 results \citep{Planck13} $\omm=0.307$, $\omb=0.048$, h=0.678, $\sigma_8=0.823$.
This allows us to compare the effects of the different cosmological parameters on the formation of first galaxies.
In particular, the value of $\omm$ in the Planck cosmology is significantly higher than the WMAP value, commonly used in previous simulations.  

We employ a standard zoom-in technique \citep{Klypin02} to generate the initial conditions.
For each set of cosmological parameters, we have three different cosmological boxes of 10, 20 and 40 $\hmpc$.
For each box, we first run a low-resolution (128$^3$ particles) N-body-only simulation ($z_{\rm init}=150$) with the \textsc{ART} code
\citep{Kravtsov97}. 
We select all distinct halos with a maximum circular velocity at $z=5$ greater than a specified threshold $V_{\rm cut}$  (\tab{1}). 
This restriction allows us to avoid including poorly-resolved halos in our sample.
The main sample consists of 978 halos.

 \begin{table} 
\caption{The FirstLight sample. The units for box size and $V_{\rm cut}$ are $\hmpc$ and $\kms$ respectively.  }
 \begin{center} 
 \begin{tabular}{ccccc} \hline 
\multicolumn{2}{c} {Box size } \ \ \ Cosmology  & Effective resolution & log($V_{\rm cut}$) & \# of halos \\
\hline 
10   & WMAP & 2048$^3$ & 1.7 &  201 \\
20  & WMAP & 4096$^3$ & 2.0 &  114 \\
40  & WMAP & 8192$^3$ & 2.3 & 31 \\
10  & Planck & 2048$^3$ & 1.7 &  344 \\
20  & Planck & 4096$^3$ & 2.0 &  228 \\
40  & Planck & 8192$^3$ & 2.3 & 60 \\ 
\end{tabular} 
 \end{center} 
\label{tab:1} 
 \end{table} 
 
  \begin{figure}
	\includegraphics[width=\columnwidth]{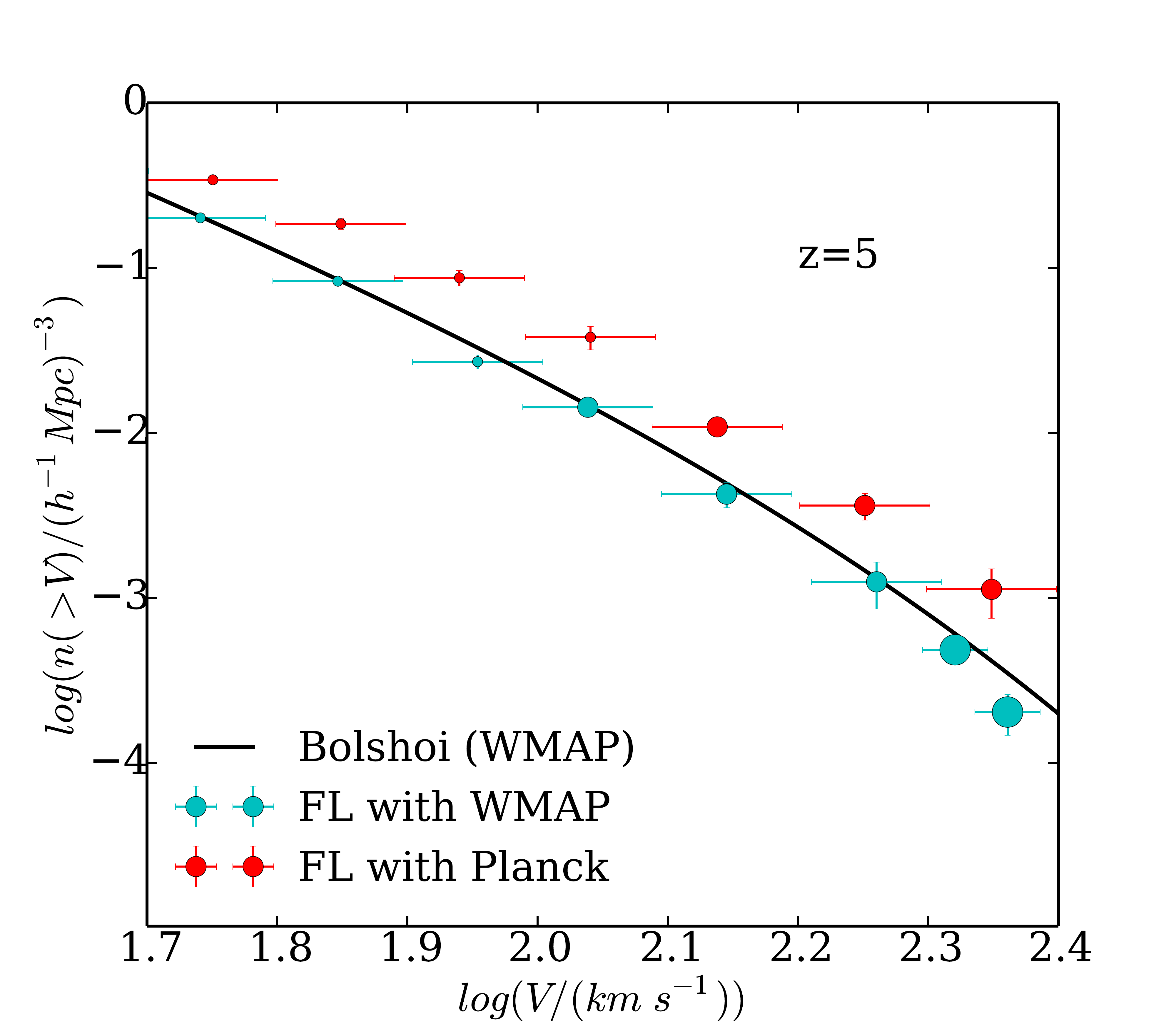}
    \caption{Velocity function of DM halos from the FirstLight sample at z=5. The size of the points increases with  box size (\tab{1}). 
    The results with a WMAP cosmology are consistent with the halo statistics from the Bolshoi simulation \citep{Klypin11} using the same cosmological parameters. By contrast, the halo number densities with Planck cosmology are significantly higher due to the higher $\omm$ value.  }
    \label{fig:Vf}
\end{figure}
 
 Using these halos we generate the velocity function at $z=5$ (\Fig{Vf}) and compare it with the results coming from large N-body-only simulations, such as the Bolshoi simulation \citep{Klypin11} with a box size of 250 $\hmpc$. 
 The halo statistics with WMAP cosmology is remarkably similar to Bolshoi, which uses the same cosmological parameters.
 This demonstrates that the halo number densities are accurately reproduced within the selected mass range, despite the relatively small cosmological volumes and the low resolution used in the selection of halos. 
 Higher resolution runs with $256^3$ particles confirm that these results have converged.
 
 The halo statistics with Planck cosmology shows significantly higher halo number densities.
 This is a consequence of the higher $\omm$ from Planck with respect to WMAP.
 In a more massive Universe, structures collapse earlier and they are denser.
  For example, the number density of distinct halos at $z\simeq7$ increases by a factor of 1.5-3, depending on their mass \citep{RodriguezPuebla}.
  This could have important implications for the formation of the first galaxies and  reionisation.

 Once the halos are selected, the initial conditions with higher resolution are generated
 using  \textsc{pmgalaxy}  \citep{Klypin11} for the 10 and 20 $\hmpc$ boxes and   \textsc{music} \citep{Music} for the 40 $\hmpc$ box.
 All three volumes have the same DM particle mass resolution of m$_{\rm DM}=10^4 \ \msun$   (a maximum effective resolution of $8192^3$ particles).
 The minimum mass of star particles is $100 \ \msun$.
 The maximum spatial resolution is always between 8.7 and 17 proper pc (a comoving resolution of 109 pc after $z=11$).
 This mass resolution is a factor of 3 better than in the Renaissance simulations \citep{OShea15} and comparable to the resolution in the FiBY project \citep{Paardekooper13,Cullen17} but in a much larger volume.
  
 \subsection{ART}
 
 The simulations are performed with the  \textsc{ART} code
\citep{Kravtsov97,Kravtsov03}, which accurately follows the evolution of a
gravitating N-body system and the Eulerian gas dynamics using an Adaptive Mesh Refinement (AMR) approach.
Besides gravity and hydrodynamics, the code incorporates 
many of the astrophysical processes relevant for galaxy formation.  
These processes, representing subgrid 
physics, include gas cooling due to atomic hydrogen and helium, metal and molecular 
hydrogen cooling, photoionisation heating by a constant cosmological UV background with partial 
self-shielding, star formation and feedback, as described in 
\citet{Ceverino09}, \citet{CDB}, and \citet{Ceverino14}. 

 In addition to thermal-energy feedback, the simulations use radiative feedback,
 as a local approximation of radiation pressure.
This model adds non-thermal pressure to the total gas pressure in regions where ionising photons from massive stars are produced and trapped. 
In the current implementation, named RadPre\_IR in \citet{Ceverino14}, radiation pressure is included in the cells (and their 6 closest neighbours) that contain stellar particles younger than 5 Myr and whose gas column density exceeds $10^{21}\ \cms.$ 
Finally, the model also includes a  moderate trapping of infrared photons, only if the gas density in the host cell exceeds a threshold of 300 $\cmc$.
More details can be found in \citet{Ceverino14}.

% Momentum injection from SN shells:

In addition to radiative feedback, the latest model also includes the injection of momentum coming from the (unresolved) expansion of gaseous shells from supernovae and stellar winds \citep{OstrikerShetty11}.
A momentum of $3 \times 10^5 \ \msun \kms$ per massive star (i.e per star more massive than 8 $\msun$) is injected at a constant rate over 40 Myr, the lifetime of the lightest star that explodes as a core-collapsed supernovae.
The resulting specific momentum when integrated over the IMF is $3.75 \times  10^3 \kms$.
The injection of momentum is implemented in the form of a non-thermal pressure, as in \citet{Ceverino14}.
%This model is similar to the injection of momentum described in \citet{Agertz13}.

This feedback model differs from other recent implementations. 
It goes beyond the thermal-only feedback \citep{Stinson06, Stinson13,  Schaye15} and it does not shut down cooling in the star-forming regions.
It does not impose a wind solution \citep{Vogelsberger14, Hopkins14}, so that outflows are generated in a self-consistent way \citep{Ceverino16}.
Our implementation is more similar to the feedback model in \cite{Agertz15}.
Within our model, both radiative and supernovae feedback act in concert and they are equally important. The combination of early feedback from radiation and late feedback from supernovae regulates star formation within galaxies \citep{Ceverino14}.

 \subsection{FirstLight tests}
 \label{sec:FLtests}
 
 As a feasibility study for this project, we generate the initial conditions for 15 halos using the WMAP cosmology  (\tab{2}).
 They cover the full range of halo masses of the FirstLight project, from $\Mv\simeq 10^9$  to  $10^{11} \ \msun$ at $z=5$.  
These are the simulations analysed in this paper.
In order to avoid poorly resolved halos, we restrict our analysis to halos with $\Mv \geq 10^9 \ \msun$. 
They contain more than $\sim8 \times 10^4$ DM particles within the virial radius.
The runs finish at  $z=6$ when the analysis of the next section is performed, with the exception of the computationally most expensive run (FL21) that finishes at $z=8$.

  \begin{table} 
\caption{
Zoom-in simulations analysed in this paper as tests of the FirstLight project. Values are computed at $z_{\rm last}=6$ unless  otherwise stated.
The units of the maximum circular velocity, V, virial mass, $\Mv$, galaxy stellar mass, $\Ms$, and rest-frame UV magnitude, $M_{UV}$ are $\kms$, $\msun$, and mag respectively.} 
 \begin{center} 
 \begin{tabular}{cccccc} \hline 
 \multicolumn{2}{c} {ID } \ \ \ log($V$(z=5)) ) & $z_{\rm last}$ & $\Mv / 10^{10}$ & $\Ms / 10^7$  & $M_{UV}$  \\
\hline 
FL01 & 1.60 & 6 & 0.11 & 0.14 & -12.2 \\
FL02 & 1.70 & 6 & 0.22 & 0.15 & -11.9 \\
FL04 & 1.72 & 6 & 0.47 & 0.48 & -14.3 \\
FL05 & 1.90 & 6 & 0.22 & 0.30 & -13.6 \\
FL06 & 1.97 & 6 & 0.28 & 0.32 & -12.3 \\
FL08 & 1.88 & 6 & 1.7   & 4.5   & -16.7\\
FL11 & 1.96 & 6 &  2.5  & 13     & -17.1 \\
FL12 & 1.90 & 6 & 0.71 & 0.83  & -15.0 \\
FL13 & 1.84 & 6 & 0.64 & 2.9    & -15.6\\
FL14 & 1.78 & 6 & 0.30 & 0.27  & -13.5 \\
FL15 & 1.70 & 6 & 0.36 & 0.62  & -14.4 \\
FL16 & 2.19 & 6 & 6.5  & 75      & -18.9 \\
FL17 & 2.15 & 6 & 6.1  & 89      & -19.3 \\
FL19 & 2.05 & 6 & 4.4  & 66      & -19.0 \\
FL21 & 2.38 & 8 & 0.26 & 0.37  & -13.9 \\
\end{tabular} 
 \end{center} 
\label{tab:2} 
 \end{table}

%%%%%%%%%%%%%%%%%%%%%%%%%%%%%%%%%%%%%%%%%%%%%%%%%%
\section{FirstLight tests at z=6}
\label{sec:six}
%%%%%%%%%%%%%%%%%%%%%%%%%%%%%%%%%%%%%%%%%%%%%%%%%%

First we analyse the FirstLight tests at redshift $z=6$, because there are many good observational constraints at that redshift.
In order to estimate the UVLF from a set of zoom-in simulations, we need to compute the rest-frame UV magnitude associated with each major progenitor of the halos selected in section \se{FLtests} (one galaxy per zoom-in), as well as the number density of halos of the same mass as the progenitor.
Our fundamental assumption here is that each object is an unbiased representative of a sample of galaxies with similar properties.
The full FirstLight sample will further constrain  the mean and scatter around these values.

The UV continuum emission ($L_{UV}$) from a galaxy is proportional to its star formation rate (SFR) and it is independent of the galaxy history if it is measured at time-scales much longer than $\sim$$2 \times 10^7 {\rm yr}$, the typical lifetime of late-O/early-B stars that dominate the UV continuum \citep{Madau98, Kennicutt98b}. Assuming a broadband filter centred at 1500 $\AA$ with a bandpass width, $\Delta \lambda/ \lambda=0.2$, the UV luminosity becomes
\begin{equation} 
L_{UV}= \frac{\rm SFR}{\sy} 8 \times 10^{27} \rm {erg} \  \rm{s}^{-1} Hz^{-1}, 
\label{eq:LUV}
\end{equation}
where SFR is measured in a period of $3 \times 10^8 {\rm yr}$. 
Using a shorter period of $10^8 {\rm yr}$ introduces fluctuations in the UV magnitude due to the particular star-formation histories of the FirstLight tests. These fluctuations add a maximum scatter of 1 magnitude around the UV magnitude measured with a period of $3 \times 10^8 {\rm yr}$. We use the longer period because we are interested in the averaged UV luminosities, independent of their particular histories. 
The full FirstLight sample will allow us to characterise the diversity of  star-formation bursts and the variations in the UV magnitude in future work.
We assume a Salpeter IMF and
and neglect the effects of dust absorption, which should be a good approximation for the primeval galaxies considered here \citep{Bouwens16a}.
Assuming a standard conversion to AB magnitudes \citep{Bouwens08}, the UV magnitude becomes
\begin{equation} 
M_{UV} = -21.91 - 2.5 \ \rm{log} \left( \frac{L_{UV}}{2.5 \times 10^{29}  \rm {erg} \  \rm{s}^{-1} Hz^{-1}} \right) . 
\label{eq:MUV}
\end{equation}

The comoving number density is obtained from the maximum circular velocity (V) of the halo hosting each galaxy.
It can be parametrised by the following expression \citep{Klypin11} coming from a large N-body simulation:
\begin{equation} 
\Phi = A V^{-3} {\rm exp} \left( -(V/V_0)^{\alpha_H} \right) , 
\label{eq:Bolshoi}
\end{equation}
where $A= 2.8 \times 10^4 ( \Mpc / \kms )^{-3}$, $\alpha_H=1.19$, and $V_0=79 \kms$ are the parameters described in \cite{Klypin11} at $z=6$, adapted to a Planck cosmology \citep{RodriguezPuebla}. 

%\adrc{More discussion?}
%\adrc{Discussion: \Fig{Vf} shows a one-to-one correspondence between V and $\Phi$.}

\begin{figure}
	\includegraphics[width=\columnwidth]{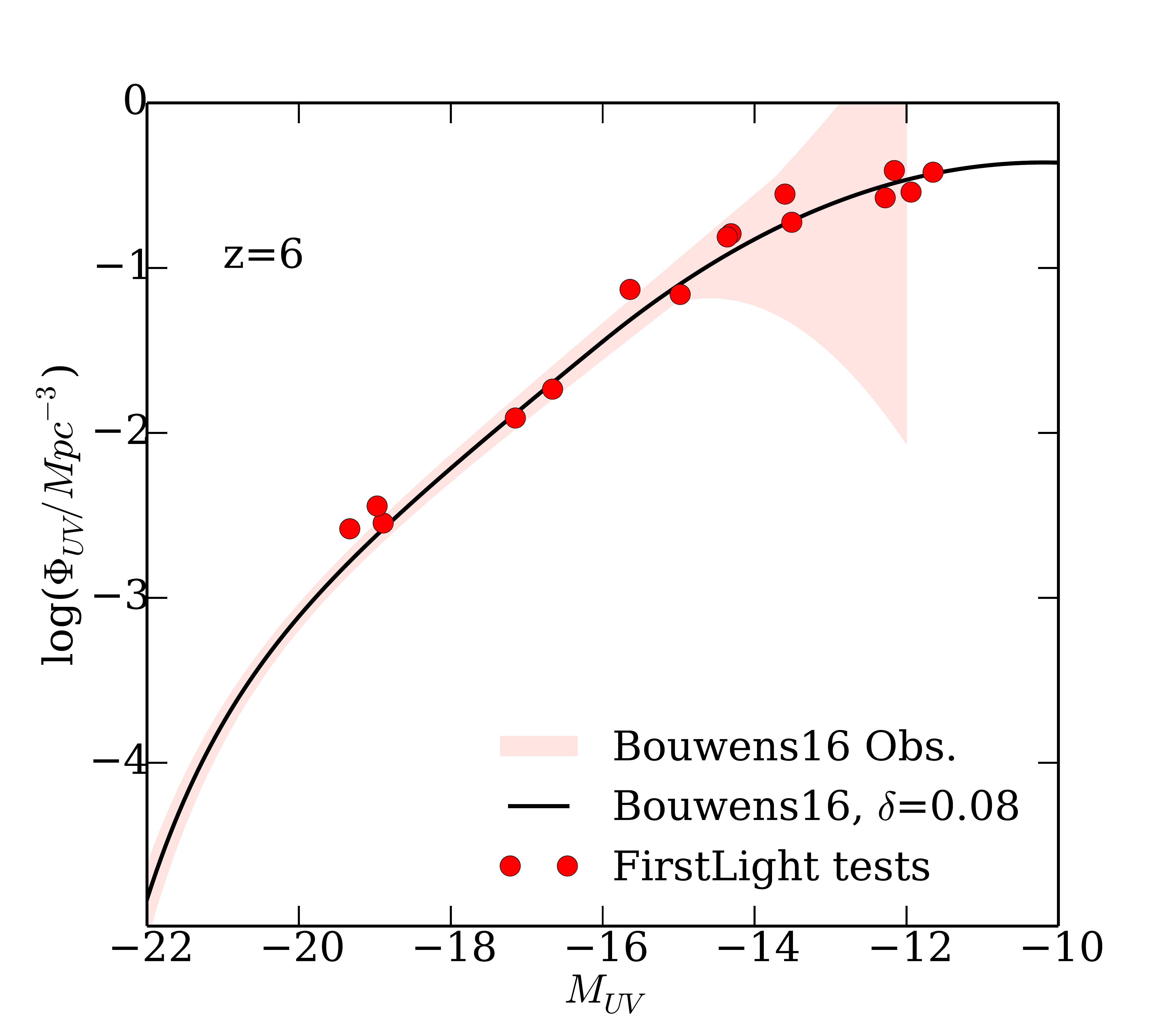}
    \caption{UV luminosity function at z=6. The FirstLight tests are consistent with observations \citep{Bouwens16}. They predict a progressive flattening of the luminosity function at low luminosities ($M_{UV}>-14$) driven by stellar feedback. }
    \label{fig:LFz6}
\end{figure}

The FirstLight luminosity function extends from high ($M_{UV}\simeq-19$) to very low ($M_{UV}\simeq-12$) luminosities (\Fig{LFz6}).
It is consistent with recent observations that are able to reach such low luminosities thanks to gravitational lensing \citep{Bouwens16}. 
In particular, the results predict a progressive flattening of the UV luminosity function at low luminosities  ($M_{UV}>-14$).
They can be parametrised by a modified version of the Schechter function \citep{Bouwens16},
\begin{equation} 
\Phi_{UV} = \Phi^* \left( \frac{ln(10)}{2.5} \right) 10^{-0.4 ( M_{UV}-M^*) (\alpha + 1) e^{-10^{-0.4 (M_{UV}-M*) } }} f ,
\label{eq:LUVfit}
\end{equation}
that includes a flattening factor for magnitudes fainter than $M_f=-16$:
\begin{equation} 
f = \left\{ \begin{array}{ll}
        10^{-0.4 \delta (M_{UV}+M_f)^2}  & \mbox{if $M_{UV} \geq M_f$};\\
        1  & \mbox{if $M_{UV} < M_f$}.\end{array} \right. 
\label{eq:f}
\end{equation}
Our value for the curvature parameter, $\delta=0.08 \pm 0.01$, is very similar to the mean parametric value used for the fiducial observations in \cite{Bouwens16}: $\delta=0.11 \pm 0.20$. The other parameters are the same as in \cite{Bouwens16}. See \tab{3}.

\begin{figure}
	\includegraphics[width=\columnwidth]{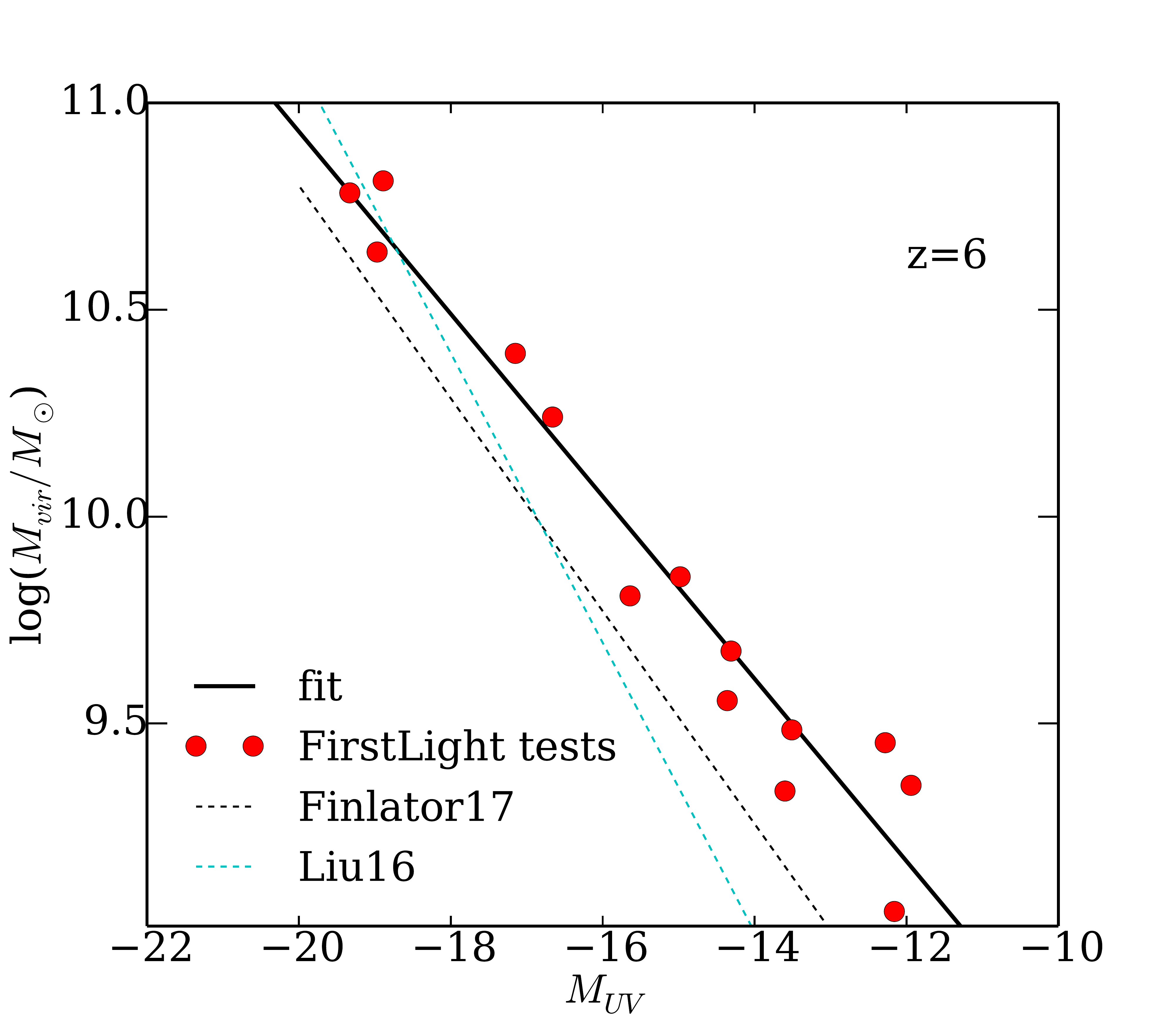}
    \caption{Virial mass versus rest-frame UV absolute magnitude at z=6. Other recent simulations \citep{Finlator17, Liu16} predict lower halo masses at a fixed UV magnitude for magnitudes fainter than $M_{UV}=-18$.  }
    \label{fig:MvMUVz6}
\end{figure}

\begin{figure}
	\includegraphics[width=\columnwidth]{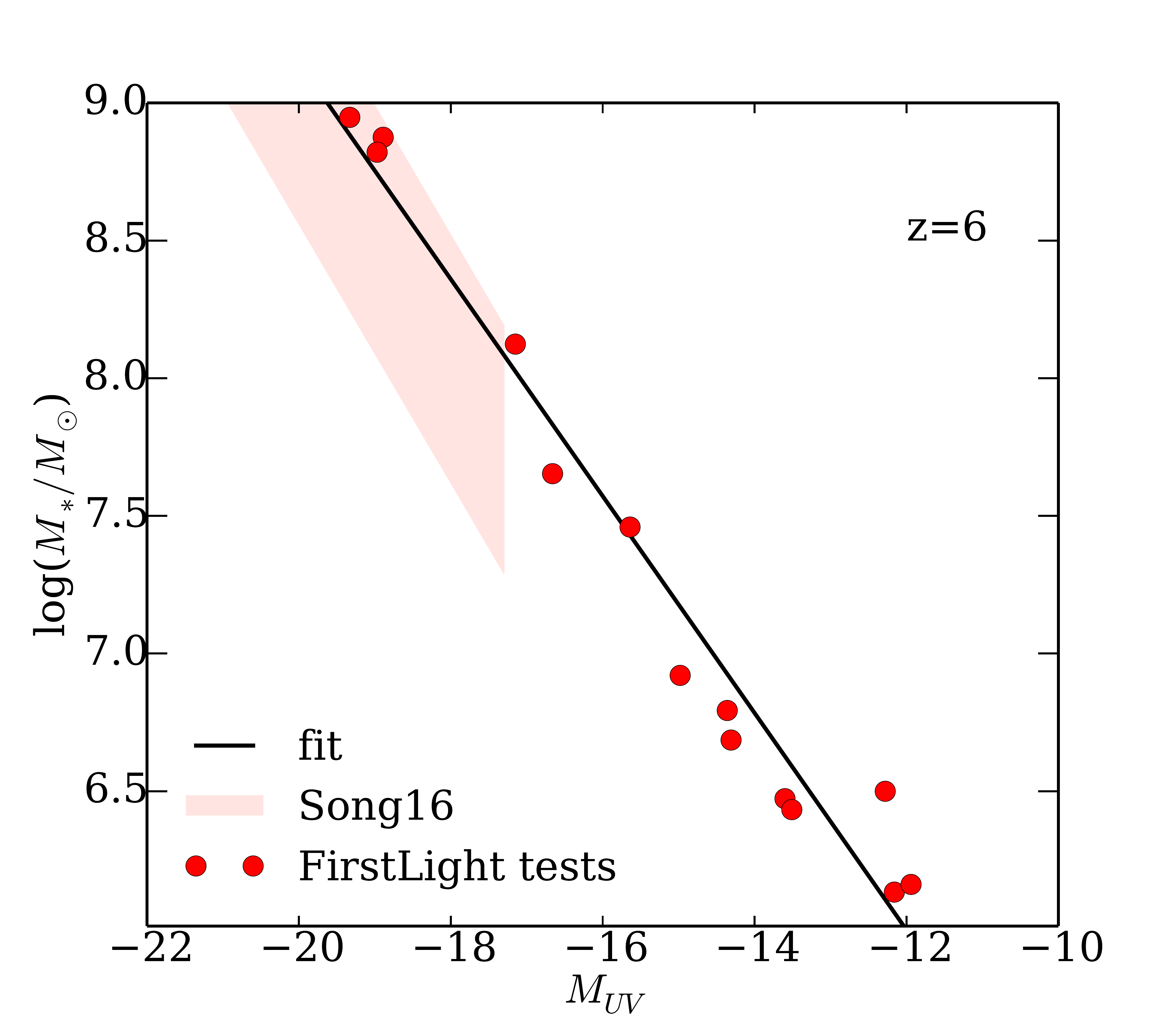}
    \caption{Stellar mass versus rest-frame UV absolute magnitude at z=6. The FirstLight tests are consistent with current observations \citep{Song16} at high luminosities  ($M_{UV}<-18$). }
    \label{fig:Starkz6}
\end{figure}

This flattening of the UVLF is produced by the progressive inefficiency of star formation at lower halo masses (higher abundances)
with maximum circular velocities of $V=30-40 \kms$.
At these masses, stellar feedback  is able to quench star formation by heating and ejecting gas that would otherwise form stars at a rate set by the cosmological gas accretion rate \citep{Dekel13}. 
Feedback is therefore able to decrease the SFR within these small galaxies, yielding UV magnitudes fainter than expected. 
This generates a flattening of the UVLF.

The above luminosity function implies a specific scaling relation between the virial mass of the hosting halo and the galaxy UV magnitude (\Fig{MvMUVz6}). For the luminosities sampled in this paper, the halo masses ranges between $\Mv \simeq 10^9$ and $10^{11} \ \msun$ at $z=6$. The relation can be fitted by the following expression,
\begin{equation} 
\left( \frac{\Mv}{10^9 \ \msun} \right) = 10^{\alpha_{\rm v} (M_{UV}-M^*_{\rm v})} ,
\end{equation}
where $ \alpha_{\rm v} = -0.2204 \pm 0.0015$, and $M^*_{ \rm v}=-11.27 \pm 0.02$.
Other recent simulations \citep{Finlator17, Liu16} predict brighter UV magnitudes for the same halo mass, especially for halos less massive than $10^{10.5} \ \msun$.
Therefore, they produce too many stars.
This is related to the different feedback models used in these simulations.
When feedback is not able to properly regulate the star formation process in low-mass halos, the simulation suffers from the overcooling problem that plagues many simulations of galaxy formation.

\begin{figure}
	\includegraphics[width=\columnwidth]{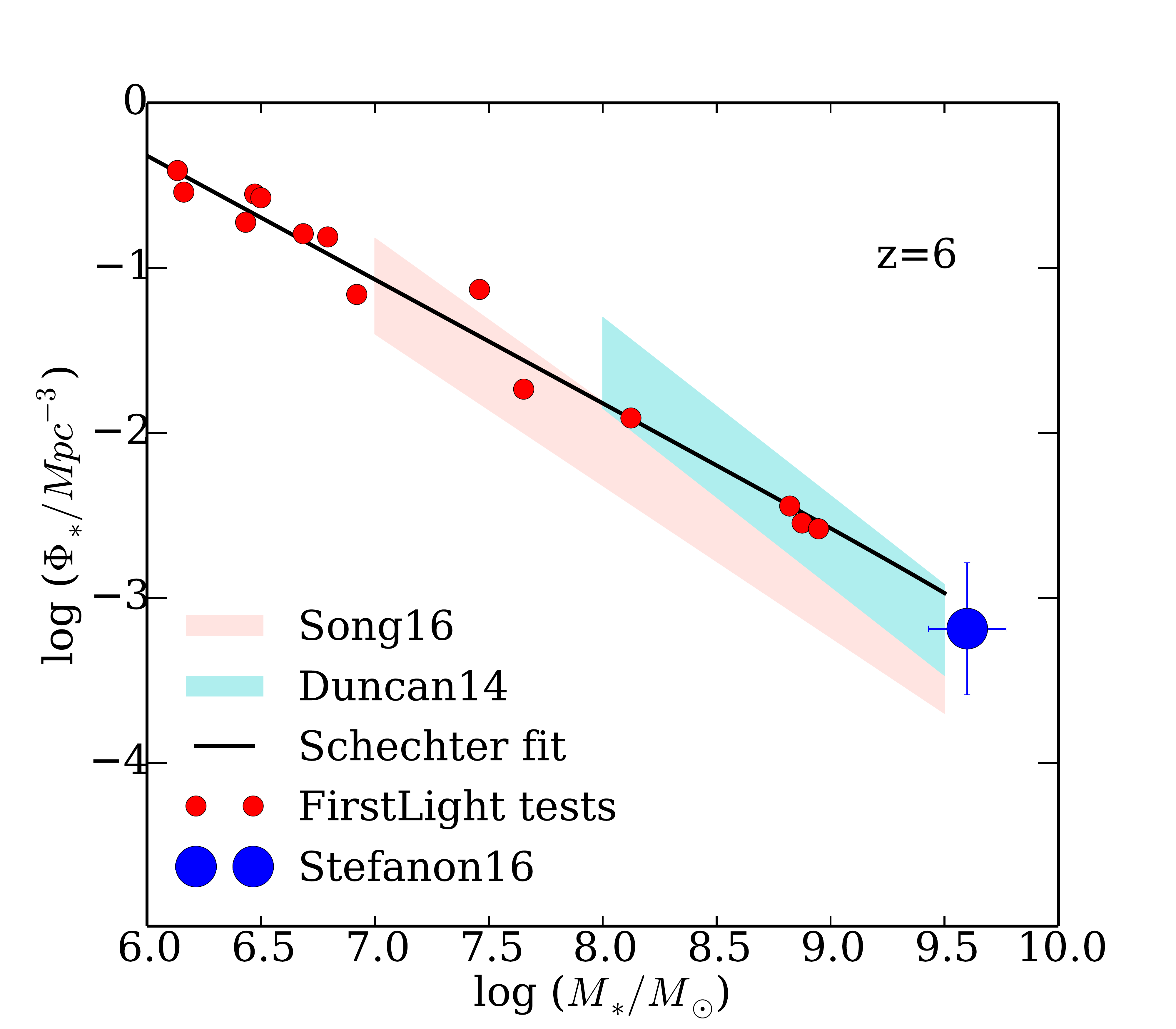}
    \caption{Stellar mass function at z=6. FirstLight results are consistent with current observations \citep{Duncan14,Song16,Stefanon16} for $\Ms>10^7 \ \msun$. For lower masses, they predict a steep mass function without signs of flattening. }
    \label{fig:MsF}
\end{figure}

The relation between stellar mass and UV magnitude is another basic scaling relation (\Fig{Starkz6}). The FirstLight results can be fitted by the following expression,
 \begin{equation} 
\left( \frac{\Ms}{10^6 \ \msun} \right) = 10^{\alpha_* (M_{UV} - M^*_*)} ,
\end{equation}
where $ \alpha_* = -0.394 \pm 0.002$, and $M^*_*=-12.13 \pm 0.03$.
The results at high luminosities  ($M_{UV}<-18$) are consistent with current observations \citep{Song16}.
The simulations contain the correct amount of stars for a given UV luminosity.
Moreover, they continue the observed trend towards lower luminosities.

%Future observations that continue the observed trends towards lower luminosities will also match the predicted relation.
%\adrc{More discussion}

The stellar mass function is also consistent with recent observations \citep{Duncan14,Song16,Stefanon16} for $\Ms>10^7 \ \msun$. The highest mass bin around $\Ms \simeq10^9 \ \msun$ has an abundance that is somewhat on the high side of the observed range constrained by \cite{Song16}. It has a better agreement with the results by \cite{Duncan14} and with observations using the luminosity function in the rest-frame visible light \citep{Stefanon16}.  
Therefore, the low-mass slope, $\alpha_s=-1.750 \pm 0.004$, is slightly shallower than the observational estimates ($-1.91 \pm 0.09$) by \cite{Song16}:
 \begin{equation} 
\Phi_* =  \Phi^*_s ln(10) 10^{(M-M^*_s)(\alpha_s+1)} e^{-10^{(M-M^*_s)}},
\label{eq:GSMF}
\end{equation}
where $M=log(\Ms$). The parameters of the Schechter fit are shown in \tab{4}. The parameter $M^*_s$ is fixed to the value reported in \cite{Song16} because the FirstLight tests do not extend to high masses and they can not constrain the exponential drop-off.
For stellar masses lower than $10^7 \ \msun$, the simulations do not show any sign of flattening of the mass function.

%%%%%%%%%%%%%%%%%%%%%%%%%%%%%%%%%%%%%%%%%%%%%%%%%%
\section{Evolution of the stellar-halo mass relation}
\label{sec:SMHM}
%%%%%%%%%%%%%%%%%%%%%%%%%%%%%%%%%%%%%%%%%%%%%%%%%%

\begin{figure}
	\includegraphics[width=\columnwidth]{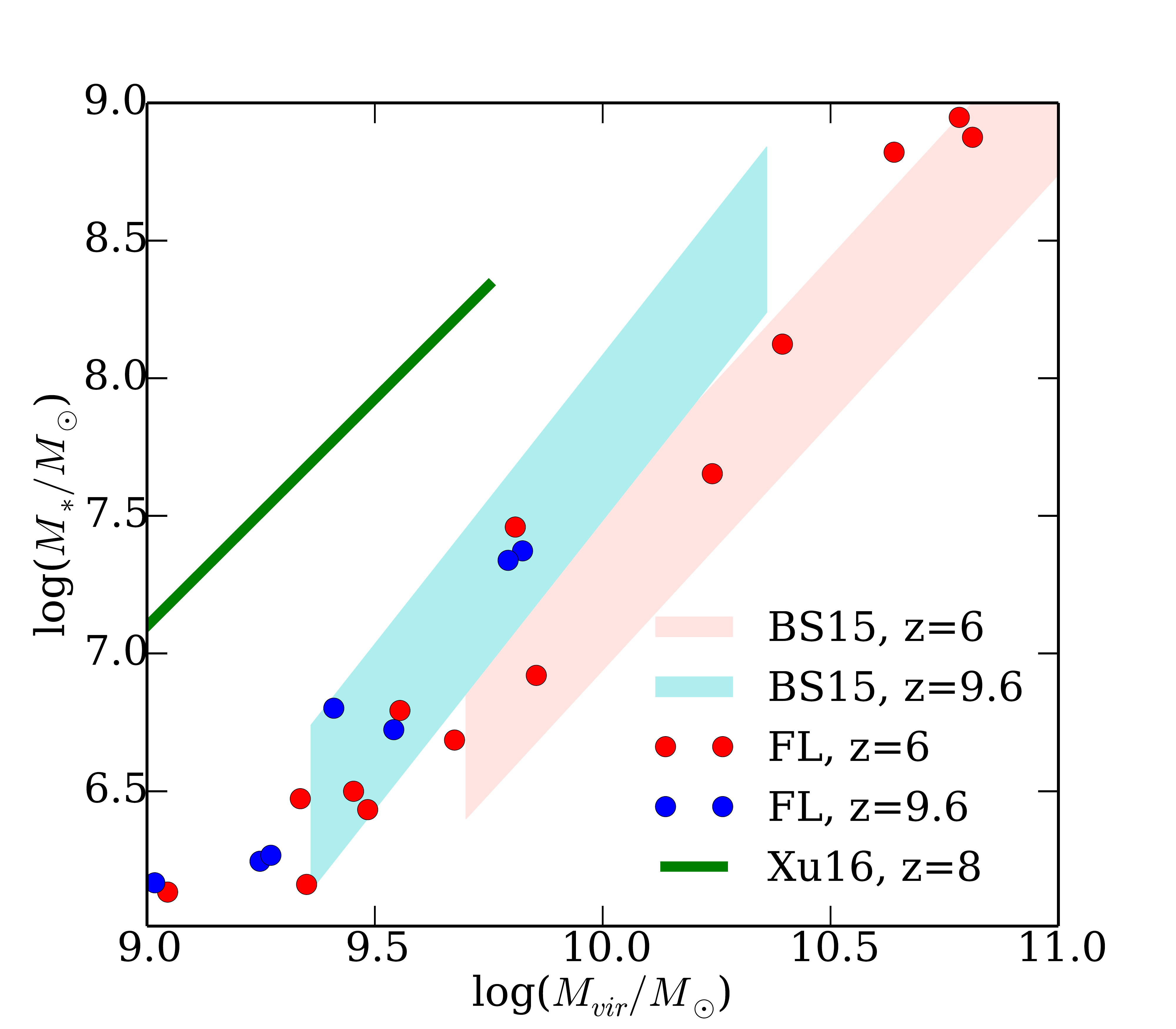}
    \caption{Evolution of the stellar-halo mass relation  between $z=9.6$ and $z=6$. The simulations are consistent with the evolution predicted by abundance matching models \citep{BehrooziSilk15}. The Renaissance simulation \citep{Xu16} forms many more stars by an order of magnitude.}
    \label{fig:SMHM}
\end{figure}

A crucial check in cosmological simulations of galaxy formation is the stellar-to-halo mass (SMHM) ratio (\Fig{SMHM}).
Model galaxies should live in halos of the right mass because many properties, such as the gas accretion rate, depend on the halo mass. However, it is difficult to measure the halo mass observationally. Therefore, we often compare with independent semi-empirical models such as abundance matching \citep{Behroozi13, Moster13}.  

After excluding a couple of outliers, the FirstLight simulations are consistent with the evolution predicted by abundance matching models \citep{BehrooziSilk15}. Between $z=6$ and 10, we found a normalisation shift of about 0.5 dex at $\Mv\simeq 10^{10} \ \msun$.
This shift is linked to differences between the evolution of the halo and galaxy mass functions.
Halos grow faster than galaxies, which are regulated by feedback. 
Therefore, galaxies of a fixed mass live in more massive halos at lower  redshifts.

This evolution does not contradict recent claims of a limited evolution in the SMHM ratio at $z\geq4$ \citep{Stefanon16} because that claim only apply to much higher masses ($\Mv\geq 10^{11.5} \ \msun$). That regime is closer to the peak of galaxy efficiency (the highest stellar-halo mass ratio), where the relation flattens and shows little evolution with redshift \citep{Behroozi13}.

Interestingly, at the lowest masses analysed in this paper, $\Mv\simeq 10^9-10^{9.5} \ \msun$, there is no evolution. This is consistent with  \cite{BehrooziSilk15} if we extrapolate their $z=6$ results to lower masses ($\Mv\simeq 10^9 \ \msun$). This is due to the fact that the relation gets steeper at higher redshifts,
mostly driven by the steepening of the halo mass function at these mass scales.

This evolution is absent in most cosmological simulations.
For example, the Renaissance simulation \citep{OShea15, Xu16} predicts a much higher stellar fraction  (\Fig{SMHM}).
Their feedback model is not efficient enought to regulate star formation. Therefore, the galaxy growth is mainly driven by the halo growth. This results in too many stars and in a time-independent SMHM ratio.

%%%%%%%%%%%%%%%%%%%%%%%%%%%%%%%%%%%%%%%%%%%%%%%%%%
\section{Evolution of the UVLF}
\label{sec:evoLF}
%%%%%%%%%%%%%%%%%%%%%%%%%%%%%%%%%%%%%%%%%%%%%%%%%%

\begin{figure}
	\includegraphics[width=\columnwidth]{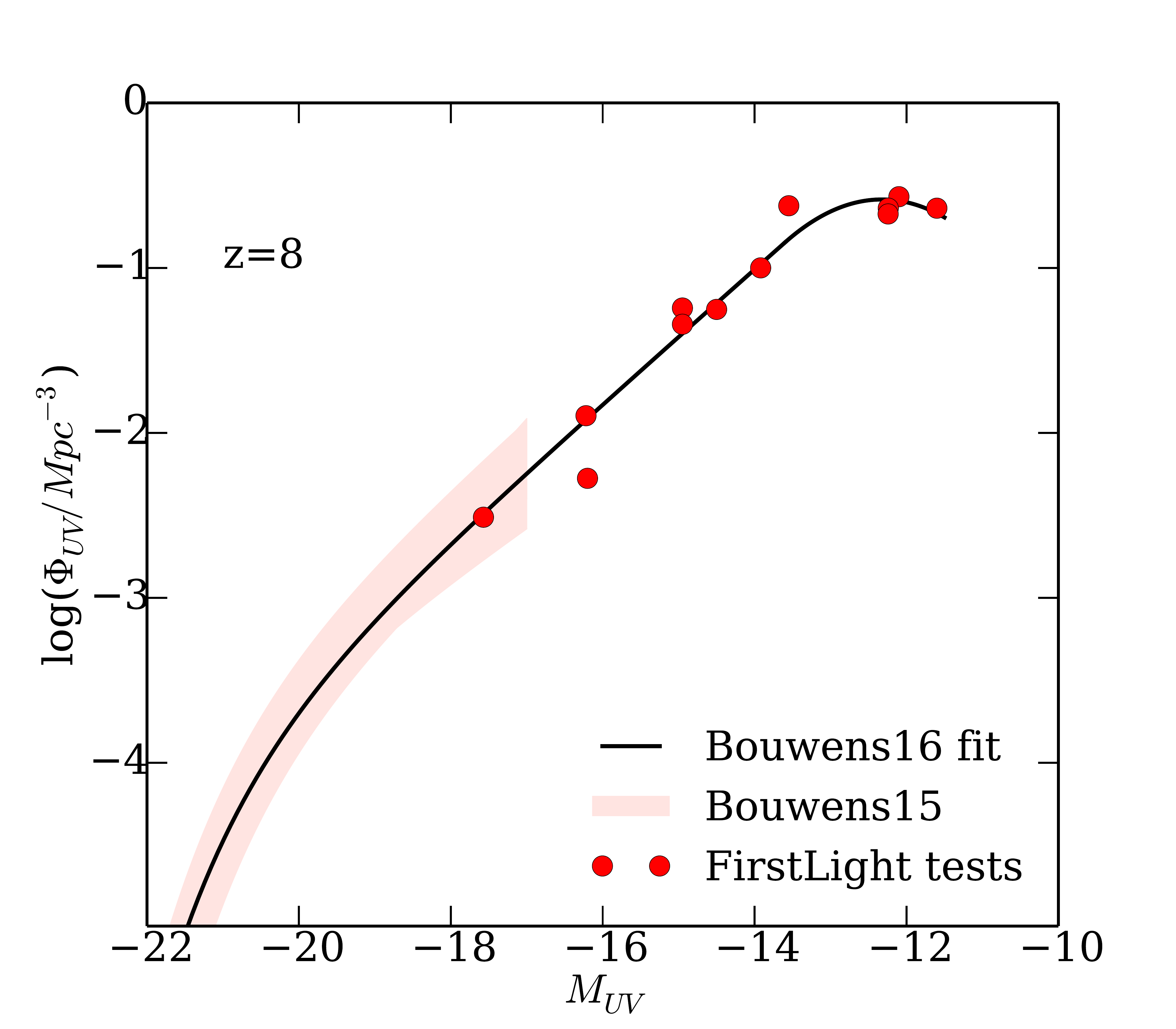}
    \caption{UV luminosity function at z=8. The FirstLight results are consistent with the extrapolation of the observed results \citep{Bouwens15} 
    with a strong flattening at low luminosities ($M_{UV}>-14$). }
    \label{fig:LFz8}
\end{figure}

  \begin{table} 
\caption{Best parameters of the Schechter fit to the simulated UVLF. Fixed values come from Bouwens et al. (2016a). }
 \begin{center} 
 \begin{tabular}{cccc} \hline 
 \multicolumn{2}{c} {z } \ \ \ \ \ \ \ \ \  $\alpha$ \ \ \ \ \ \ \ \ \ \ \ & $M^*$  & $\Phi^*/ 10^{-3}  \Mpc^{-3}$    \\
\hline 
6   & -1.92 & -20.94 & 0.57 \\
8   & -2.02 & -20.63 & 0.21 \\
10 & $-2.65 \pm 0.15$ & -20.92 & $0.008 \pm 0.005$ \\
\end{tabular} 
 \end{center} 
\label{tab:3} 
 \end{table} 

\Fig{LFz8} shows the UVLF at $z=8$.
The FirstLight results extend the observed function \citep{Bouwens15} towards lower luminosities with some overlap at $M_{UV}\simeq-18$.
The fit given in \cite{Bouwens16}, \Equ{LUVfit}, provides an excellent description using the same parameters (\tab{3}).
At low luminosities   ($M_{UV}>-14$) a flattening of the UVLF is clearly visible. This can be parametrised by \Equ{f} using $M_f=-13.6 \pm 0.2$ and $\delta=0.4 \pm 0.1$. 
This flattening is more pronounced than at $z=6$ (high $\delta$), partially due to the increase in the slope of the UVLF at higher redshifts (higher $\alpha$).

\begin{figure}
	\includegraphics[width=\columnwidth]{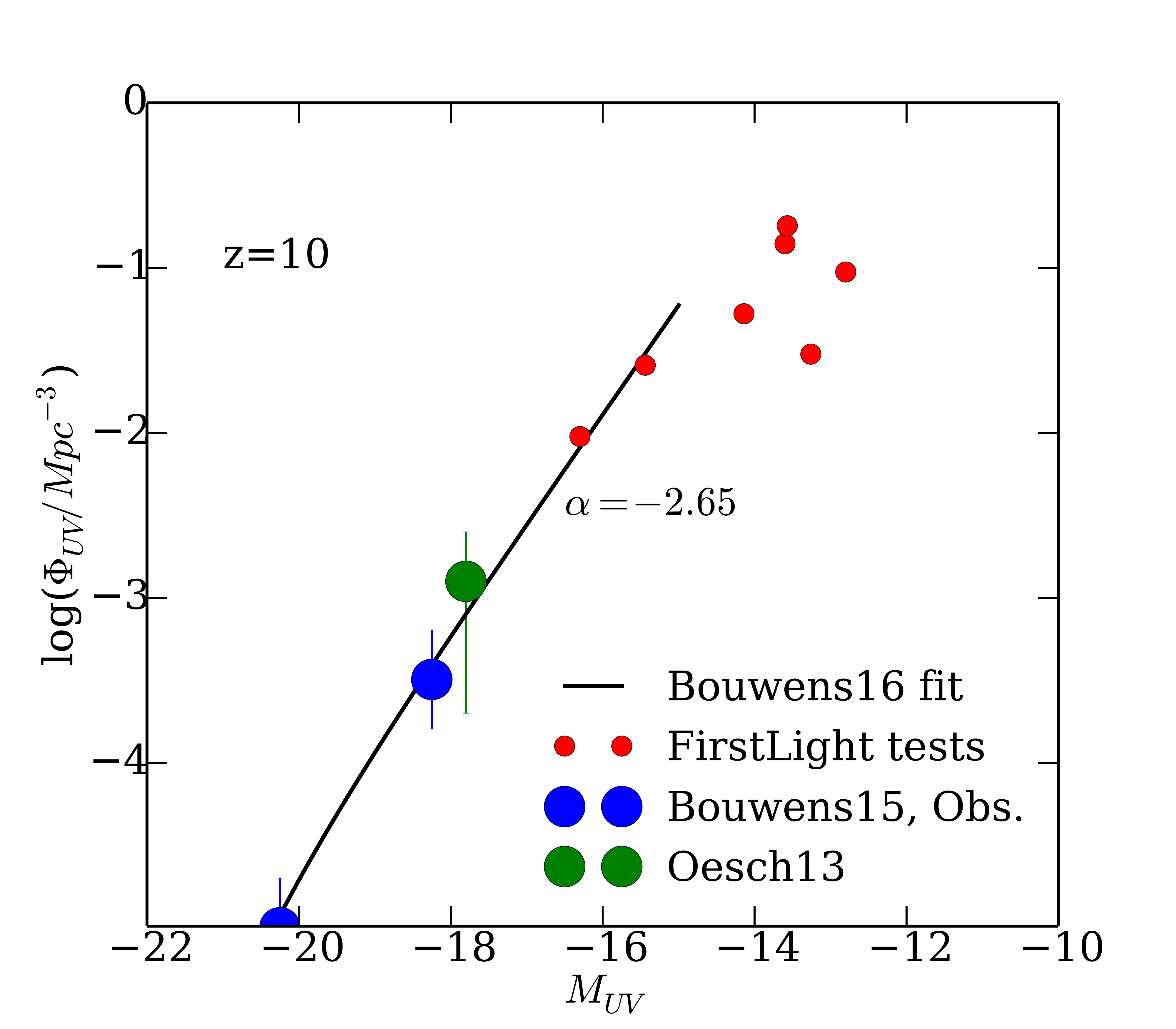}
    \caption{UV luminosity function at z=10. FirstLight and observations \citep{Oesch13,Bouwens15} give a very steep power-law slope with $\alpha=-2.65$. }
    \label{fig:LFz10}
\end{figure}

\Fig{LFz10} shows the UVLF at $z=10$ from FirstLight and observations \citep{Oesch13,Bouwens15}.
All combined results predict a very steep power-law slope for magnitudes brighter than -15 ($\alpha=-2.65 \pm 0.15$).
This value is higher than the expected value of -2.27 based on extrapolations of observations at lower redshifts \citep{Bouwens15}. 
We predict a faster evolution of the LF slope at $z\geq10$, in agreement with simple semi-analytical models \citep{Mason15} that also predict a similar slope of $-2.47 \pm 0.26$. 
The main driver of this evolution is just the fast growth and assembly of halos at these high redshifts.

Due to the large scatter of points at low luminosities ($M_{UV} > -14$), we cannot constrain the flattening of the LF at this redshift.
This part of the parameter range will be better covered with the full FirstLight survey.

%%%%%%%%%%%%%%%%%%%%%%%%%%%%%%%%%%%%%%%%%%%%%%%%%%
\section{Evolution of the stellar mass function}
\label{sec:evoMs}
%%%%%%%%%%%%%%%%%%%%%%%%%%%%%%%%%%%%%%%%%%%%%%%%%%

\begin{figure}
	\includegraphics[width=\columnwidth]{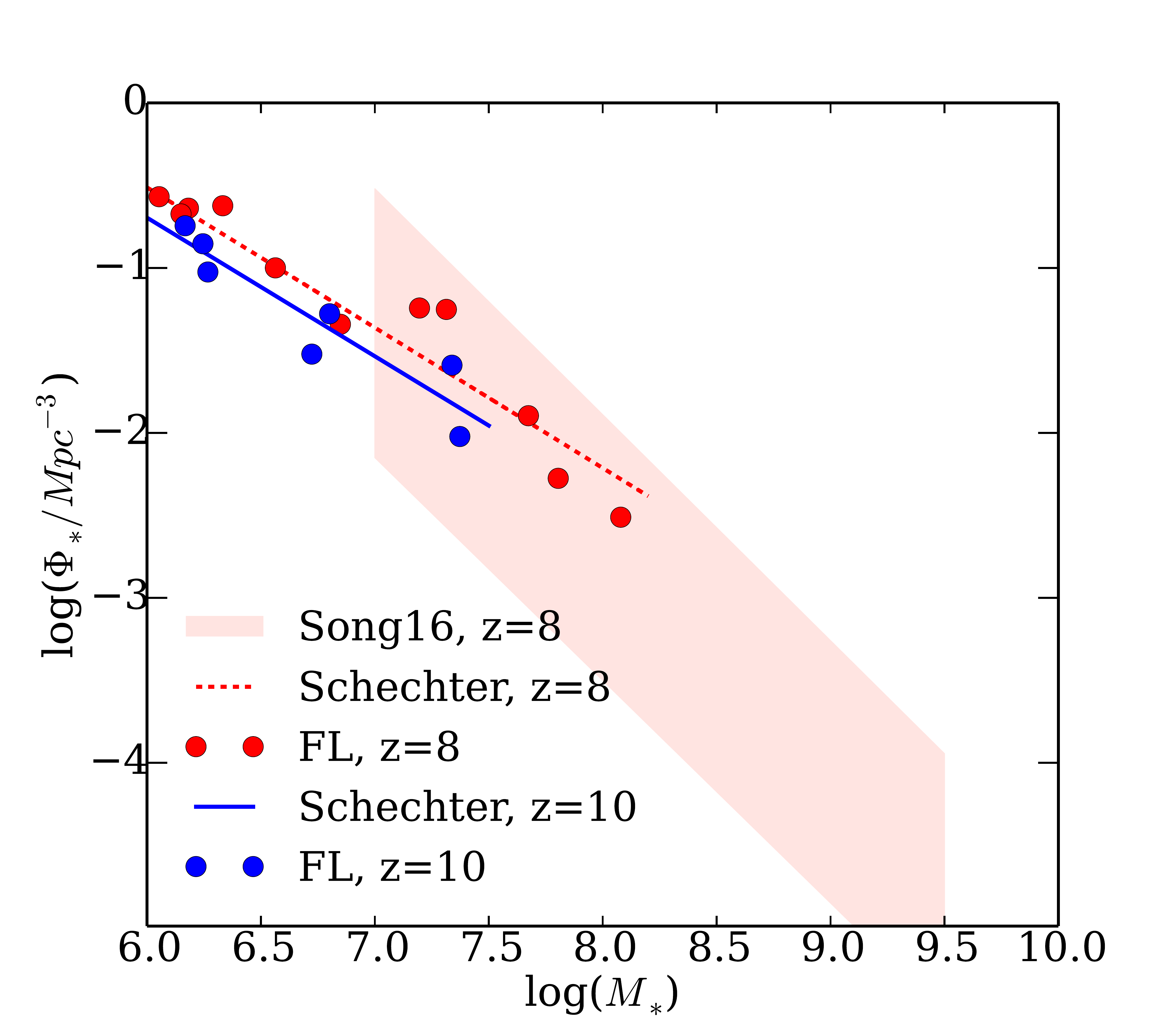}
    \caption{Stellar mass function at z=8 and $z=10$. FirstLight results are consistent with current observations at $z=8$ \citep{Song16}, although the slope is slightly lower that the observational estimates, $\alpha_s=-1.85$. The mass function at $z=10$ shows a similar slope and a slightly lower normalisation.}
    \label{fig:MsF8_10}
\end{figure}

  \begin{table} 
\caption{Best parameters of the Schechter fit to the simulated galaxy stellar mass function. 
Fixed values come from Bouwens et al. (2016a).}
 \begin{center} 
 \begin{tabular}{cccc} \hline 
 \multicolumn{2}{c} {z }  \ \ \ \ \ \ \ \ \ $\alpha_s$ \ \ \ \ \ \ \ & $M^*_s$  & $\Phi^*_s/ 10^{-5}  \Mpc^{-3}$    \\
\hline 
6   & $-1.75 \pm 0.04$ & 10.72 & $ 6.0 \pm 0.4$ \\
8   & $-1.85 \pm 0.04$ & 10.72 & $1.3 \pm 0.4$ \\
10 & $-1.84 \pm 0.12$ & 10.72 & $0.95 \pm 1$ \\
\end{tabular} 
 \end{center} 
\label{tab:4} 
 \end{table} 

\Fig{MsF8_10} shows the stellar mass function at $z=8$. 
The FirstLight tests are consistent with observations \citep{Song16} for stellar masses higher than $\Ms>10^7 \ \msun$. 
The simulations extend to lower masses, $\Ms=10^6 \ \msun$, and the slope, $\alpha_s=-1.85 \pm 0.04$,
is close to the lower limit of
the observational constraints, $-2.2 \pm 0.5$ \citep{Song16}.
The simulated sample lacks galaxies in the high-mass regime, $\Ms>10^8 \ \msun$ at this redshift. 
Therefore, we cannot exclude a steeper slope at high masses plus a different slope below a stellar mass of $10^{6.5} \ \msun$. 
Future simulations will clarify this issue.

Finally, \Fig{MsF8_10} also shows the stellar mass function at $z=10$. 
At this redshift, there are no observational estimates. We can only predict the mass function in the low mass regime, between $10^6$ and $10^{7.5} \ \msun$. The low-mass slope does not evolve much between these redshifts, although the normalisation is significantly lower at $z=10$ (\tab{4}).

%%%%%%%%%%%%%%%%%%%%%%%%%%%%%%%%%%%%%%%%%%%%%%%%%%
\section{Conclusions and Discussion}
\label{sec:conclusions}
%%%%%%%%%%%%%%%%%%%%%%%%%%%%%%%%%%%%%%%%%%%%%%%%%%

We have introduced the FirstLight project that aims to generate a large database of simulated galaxies around the epoch of reionisation ($z\geq6$), with an unprecedented numerical resolution (an effective resolution of upto $8192^3$ particles).
The first tests of this program, a set of 15 zoom-in, cosmological simulations, yield the following main results:
\begin{itemize}
\item
The simulations agree well with the best observational constraints at $z=6$, such as the UVLF, the stellar mass-UV magnitude relation, and the galaxy stellar mass function.
\item
 The UVLF starts to flatten below $M_{UV}>-14$ for halos with maximum circular velocities of $V=30-40 \kms$. This flattening is due to stellar feedback.
\item
The stellar-halo mass relation evolves from $z=6$ to $z=10$ according to the expectations from abundance matching models \citep{BehrooziSilk15}.
\item
The power-law slope of the UVLF evolves rapidly with redshift, reaching a value of $\alpha\simeq-2.5$ at $z=10$.
\item
On the other hand, the galaxy stellar mass function evolves slowly with time between $z=8-10$, in particular at the low-mass end.
\end{itemize}

% Final discussion:

The FirstLight project satisfies the need for a large sample of zoom-in calculations with
high predictive power for the astrophysical interpretation of the expected wealth of  data from new facilities like JWST, WFIRST and 30-meters-class telescopes. A future mock survey of synthetic observations can be directly compared with current and future surveys.

Thanks to the large number statistics, the full FirstLight sample is able to address the mean galaxy properties over a large range of halo masses. It can shed light on the physical origin of the galaxy scaling relations and their evolution during the early galaxy assembly. 

The shape of galaxies at high redshifts is very different from local counterparts. 
They tend to be clumpy, irregular or even elongated \citep{Ceverino15b}.
The simulated galaxies will be well resolved, and therefore the mock survey will cover a large diversity of galaxy morphologies. This project will uncover the key mechanisms of morphological transformation, in relation with galaxy efficiency and star-formation self-regulation by feedback.

%Caveats & missing physics:

Many physical processes are missing in the current simulations: non-equilibrium cooling,  local photoionisation and photoheating, radiative transfer effects, population-III or black-hole physics.
They are all important in different regimes and situations. Future simulations using the same initial conditions will include some of these effects.
However, based on the good agreement between the global properties of the simulated galaxies and current observational constraints, the above physical processes do not seem crucial for the formation of galaxies within the mass and redshift range explored in this paper. 

%\adrc{feedback driven flattening of the UVLF. discussion on filtering mass and photoevaporation scale and other effects towards lower V. fraction of dark halos. See OShea15.}

%%%%%%%%%%%%%%%%%%%%%%%%%%%%%%%%%%%%%%%%%%%%%%%%%%
\section*{Acknowledgements}
%%%%%%%%%%%%%%%%%%%%%%%%%%%%%%%%%%%%%%%%%%%%%%%%%%

We thank Peter Behroozi, Joop Schaye, and Sandro Tacchella for fruitful discussions. This work has been funded by  the ERC Advanced Grant, STARLIGHT: Formation of the First Stars (project number 339177).
The authors gratefully acknowledge the Gauss Centre for Supercomputing for funding this project by providing computing time on the GCS Supercomputer SuperMUC at Leibniz Supercomputing Centre.
The authors acknowledge support by the state of Baden-Württemberg through bwHPC
and the German Research Foundation (DFG) through grant INST 35/1134-1 FUGG.

%%%%%%%%%%%%%%%%%%%%%%%%%%%%%%%%%%%%%%%%%%%%%%%%%%



\bibliographystyle{mnras}
\bibliography{FirstLight4} % if your bibtex file is called example.bib

%%%%%%%%%%%%%%%%%%%%%%%%%%%%%%%%%%%%%%%%%%%%%%%%%%

% Don't change these lines
\bsp	% typesetting comment
\label{lastpage}
\end{document}

% End of mnras_template.tex